\documentclass[11pt]{article} 
\usepackage{amssymb}     
\usepackage{latexsym}    
\usepackage{epsfig}  
\usepackage{diagrams} 
\def\endproof{\hfill$\Box$} 

\newtheorem{Th}{Theorem}[section] 
\newtheorem{theorem}[Th]{Theorem}
\newtheorem{proposition}[Th]{Proposition}      
\newtheorem{lemma}[Th]{Lemma}
\newtheorem{corollary}[Th]{Corollary}
\newtheorem{definition}[Th]{Definition}  

\newtheorem{remark}[Th]{Remark}
\newtheorem{problem}[Th]{Problem}

\newcommand{\bit}{\begin{itemize}}
\newcommand{\eit}{\end{itemize}\par\noindent}
\newcommand{\ben}{\begin{enumerate}}
\newcommand{\een}{\end{enumerate}\par\noindent}
\newcommand{\beq}{\begin{equation}}
\newcommand{\eeq}{\end{equation}\par\noindent}
\newcommand{\beqa}{\begin{eqnarray*}}
\newcommand{\eeqa}{\end{eqnarray*}\par\noindent} 
\newcommand{\beqn}{\begin{eqnarray}}  
\newcommand{\eeqn}{\end{eqnarray}\par\noindent}          

\title{\bf Entropic Geometry from Logic}
\author{Bob Coecke \\ 
{\small Oxford University Computing Laboratory}\\
{\footnotesize Wolfson Building, Parks Road, Oxford OX1 3QD, UK.}\,;
{\footnotesize {Bob.Coecke@comlab.ox.ac.uk}}  
}
\date{} 
\begin{document}   
\maketitle 

\vspace{-0mm}\noindent
{\bf Abstract.} 
We establish the following equation:
\vspace{-1.8mm}
\noindent
\begin{center}
Quantitative Probability = Logic + Partiality of Knowledge + Entropy
\end{center}
\vspace{-1.8mm}
\noindent
That is:~1.~A finitary probability space $\Delta^n$ (=\,all probability
measures on $\{1,\ldots,n\}$)
can be fully and faithfully represented by the pair consisting of the
abstraction $D^n$ (=\,the object up to isomorphism) of the partially ordered 
set $(\Delta^n,\sqsubseteq)$ introduced in \cite{CoeMar}, and, Shannon entropy;
2.~$D^n$ itself can be obtained 
via a systematic purely order-theoretic
procedure (which embodies introduction
of partiality of knowledge) on
an (algebraic) logic.  This procedure applies to any poset $A$; 
$D_A\cong(\Delta^n,\sqsubseteq)$ when
$A$ is the $n$-element powerset
%%%%%%%%%%%%%
%:={\rm P}(\{1,\ldots,n\})$ 
%%%%%%%%%%%%%
and $D_A\cong(\Omega^n,\sqsubseteq)$,
the domain of mixed quantum states also introduced in \cite{CoeMar}, when
$A$ is
%%%%%%%%%%%%%
%:=\mathbb{L}^n$, 
%%%%%%%%%%%%%
the lattice
of subspaces of a Hilbert space.

\section{Introduction}

For a century the dominant formalization of uncertainty has been in
terms of measures on a support.  However, already in 1926 F.~P.~Ramsey
proposed to conceive probability as {\it the logic of partial
knowledge\,} \cite{Ramsey}. 
D.~S.~Scott relied on a more general notion of {partiality\,} to
propose the mathematical structure
of a {\it domain\,}
\cite{Scott}.  A deep connection between domains and
measures of content was established by K. Martin in \cite{Martin}.  A
domain $(\Delta^n,\sqsubseteq)$ of
probability measures which has Shannon entropty as a measure of content
and a domain
$(\Omega^n,\sqsubseteq)$ of mixed quantum states which has von Neumann
entropy as a measure of content
were introduced in
\cite{CoeMar}.  In this paper, we establish:
\ben
\item Quantitative Probability = Qualitative Probability + Entropy
\item Qualitative Probability := Logic + Partiality of Knowledge 
\een
The first claim follows from the fact
that $D^n$, the abstraction of
${(\Delta^n,\sqsubseteq)}$ as a partially ordered set (=\,{\it poset}),
when equipped with Shannon
entropy $\mu$, fully and faithfully captures $\Delta^n$:
the identity is the only entropy-preserving order-isomorphism of
$(\Delta^n,\sqsubseteq,\mu)$ ---
up to permutation of the
names of its pure states
(Corollary
\ref{cl:iso}). Thus, uncertainty can be captured by
combining qualitative
(=\,domains) and quantitative (=\,entropy) notions of information.

A probability space does not
only admit a notion of
partiality (=\,domain structure); $D^n$ can be
purely
order-theoretically constructed in terms of partial knowledge starting
from an algebraic logic,
namely the powerset of its maximal elements. Thus, no probability space is {\it
a priori\,} required to
produce $D^n$. This establishes the second claim.
This result extends to the quantum case.  It can
be seen as the converse to 
\cite{CoeMar} Theorems 4.8 and 4.11, were the the powerset 
${\rm P}(\{1,\ldots,n\})$ and
the lattice of subspaces
$\mathbb{L}^n$ of a
$n$-dimensional Hilbert space ${\cal H}^n$ are recoverd in order-theoretic
manner respectively from
$(\Delta^n,\sqsubseteq)$ and $(\Omega^n,\sqsubseteq)$. The fact that
the {\it quantum logic\,} $\mathbb{L}^n$ constitutes the algebra of
physical properties of a quantum system \cite{BvN,CMW}, 
as opposed to the classical logic ${\rm
P}(\{1,\ldots,n\})$, justifies the utterance {\it probability from
logic\,} (Section \ref{sec:const}).

In fact, we
produce a probability space with, {\it in addition}, a partial order
relation on it (so the above equations are understatements).
(Pre)orders have been in the study of probability \cite{Muir}, but never
captured probability itself.
 
\section{Preliminaries}
 
In this section we recall results from \cite{CoeMar}. Let $\Delta^n$ be all
probability distributions on
$\{1,\ldots,n\}$, that is, either a list 
$x=(x_1,\ldots,x_n)\in[0,1]^n$ or a map
$x:\{1,\ldots,n\}\to[0,1]::i\mapsto x_i$, with  
$\sum_{i=1}^{i=n}x_i=1$.
Decreasing monotone distributions in $\Delta^n$, i.e., for all
$i\in\{1,\ldots, n-1\}$ we have
$x_i\geq x_{i+1}$, are denoted by
$\Lambda^n$.    
The {\it spectrum} of $x$ is the set
${\rm spec}(x):=\{x_i\mid 1\leq i\leq n\}$.
Denote the collection of all permutations 
$\sigma:\{1,\ldots,n\}\to\{1,\ldots,n\}$ as $S(n)$. For a poset $D$, 
we set $\uparrow\!x=\{y\in D\mid x\sqsubseteq y\}$ and 
$\downarrow\!x=\{y\in D\mid y\sqsubseteq x\}$;
we call $e\in D$ {\it maximal} iff $\uparrow\!e=\{e\}$; we denote the
set of maximal elements of $D$ by
${\rm Max}(D)$; the {\it bottom} $\bot$ (if it exists) of $D$ is
defined by
$\uparrow\!\bot=D$.  A poset $D$ is a
{\it chain\,} 
iff $x,y\in D$ either implies
$x\sqsubseteq y$ or $y\sqsubseteq x$.
\begin{definition}
\label{classicalsymmetries}
Let $n\geq 2$.  For $x,y\in\Delta^n$, we have $x\sqsubseteq y$  iff
there exists $\sigma\in S(n)$ such that
$x\cdot\sigma, y\cdot\sigma\in\Lambda^n$ 
and if we have $\forall i\in\{1,\ldots,n-1\}$:
\beq\label{eq:monosym}
(x\cdot\sigma)_i(y\cdot\sigma)_{i+1}\leq
(x\cdot\sigma)_{i+1}(y\cdot\sigma)_i\,.
\eeq
\end{definition}   
\begin{theorem}
Let $n\geq 2$. Then, $(\Delta^n,\sqsubseteq)$ is a partially ordered set
with 
\[
{\rm Max}(\Delta^n)=\left\{e\in \Delta^n\mid {\rm
spec}(x)=\{0,1\}\right\}
\quad\&\quad
\bot=(1/n,\ldots,1/n)\,.
\]
Moreover, it is a dcpo and admits the notions of partiality
and approximation, that is, $(\Delta^n,\sqsubseteq)$ is entitled to be called a 
domain.\footnote{We refer to \cite{CoeMar} for definitions
and details on these domain-theoretic aspects.  They are not essential
for the developments in this paper.}
Finally, Shannon entropy 
\[
\mu:\Delta^n\to [0,1]::x\mapsto-\sum_{i=1}^nx_i\log x_i
\]
is a
measure of content in the sense of  \em
\,\cite{Martin}\em.\footnote{I.e., there is a tight 
connection between $\mu$ and the
domain-theoretic properties of $(\Delta^n,\sqsubseteq)$.}
\end{theorem}
The intuition behind $x\sqsubseteq y$ is:  
``{\it State\,} $y$ is more {\it informative\,}
than {\it state\,}
$x$''. 
In epistemic terms this becomes: 
``{\it Observer\,}
$y$ has more {\it knowledge\,} about the system than {\it observer\,}
$x$''. Now we will formalize this intuition.  
Define the 
{\it Bayesian projections\,} $\{p_i\}_i$ such that for 
all ${x\in\Delta^{n+1}}$ with $x_i<1$:
\[
p_i(x)=\frac{1}{1-x_i}(x_1,\ldots,{x_{i-1}},{x_{i+1}},\ldots,x_{n+1})\in\Delta^n\,.
\]
We then have for $x,y\in\Delta^{n+1}$
in terms of $(\Delta^n,\sqsubseteq)$:
\beq\label{inductiverule}
x\sqsubseteq y\ \Longleftrightarrow\ (\forall i)(x_i,y_i<1\Rightarrow
p_i(x)\sqsubseteq p_i(y))\,.
\eeq
This interprets as follows. (For a detailed exposition see
\cite{CoeMar} \S2.1 and \S4.4.)~~The pure states $\{e_i\}_i$ are to be seen as
the actual states the system can
be in, while general mixed states $x$ and $y$ should be conceived as being
epistemic. 
Equivalence (\ref{inductiverule}) expresses: 1.~Whenever a state $x$ stands for
less knowledge about the system than
state $y$, then, after Bayesian update with respect to the new knowledge
that the actual state of the system
is not $e_i$, the state
$p_i(x)$ still stands for less knowledge than $p_i(y)$ due to the initial
advantage in knowledge of $y$ as
compared to
$x$; 2.~This behavior of $\sqsubseteq$ w.r.t.~knowledge update exactly
defines $\sqsubseteq\,$.\footnote{This gets extremely close to how order on physical
properties is defined \cite{JP}.}
Indeed, the {\it inductive rule\,} (\ref{inductiverule})
provides a definition equivalent to Definition \ref{classicalsymmetries} when a base
case $n=2$ is postulated as:  
\begin{definition} \label{twostates} 
For $x,y\in\Delta^2$ we set
\[
(x_1,x_2)\sqsubseteq(y_1,y_2)\ \Longleftrightarrow\ (y_1\leq x_1\leq 1/2)\mbox{
or }(1/2\leq x_1\leq y_1)\,.
\]
\end{definition}   
\begin{theorem}\label{mixing}  
The order of Definition \ref{twostates} is the only partial order on
$\Delta^2$ which has $\bot=(1/2,1/2)$ and satisfies the mixing law: 
\[
x\sqsubseteq y\ \mbox{and}\ p\in[0,1]\ \Longrightarrow\
x\sqsubseteq(1-p)x+py\sqsubseteq y\,.
\]  
\end{theorem}
The canonicity of this choice for the order on $\Delta^2$ reflects in
the shape of the Shannon entropy curve (left) and the graph of the
order (right):
\vspace{1.8mm}
\begin{center}
\hspace{-0.6cm} 
\begin{picture}(60,40)   
\qbezier(0,0)(20,60)(40,0) 
\put(0,0){\vector(1,0){50}}
\put(0,0){\vector(0,1){40}}
\put(-10,35){\normalsize$\mu$} 
\put(45,5){\normalsize$x_1$}
\end{picture}
\hspace{9mm}$\stackrel{flip}{\longrightarrow}$\hspace{9mm}
\begin{picture}(60,43)  
\qbezier(0,40)(20,-20)(40,40)
\put(-10,45){\scriptsize$(1,0)$}
\put(32,45){\scriptsize$(1,0)$}
\put(17.2,0){\scriptsize$\bot=({1\over2},{1\over2})$}
\end{picture}
\hspace{5mm}$\stackrel{straighten}{\longrightarrow}$\hspace{5mm}
\begin{picture}(40,43)  
\put(0,40){\line(3,-5){20}}      
\put(20,6.5){\line(3,5){20}}      
\put(-10,45){\scriptsize$(1,0)$}
\put(32,45){\scriptsize$(1,0)$}
\put(16.9,-1){\scriptsize$\bot=({1\over2},{1\over2})$}
\end{picture} 
\end{center}
\vspace{-0.2mm}\par\noindent  
  
Conclusively, there exists an order on $\Delta^n$
which canonically arises from envisioning probability
distributions as
informative objects, and which is tightly intertwined with Shannon entropy.  

\section{Symmetries and degeneration}\label{sec:degeneration}

For $x\in \Delta^n$, the map
$x^\Lambda:=x\cdot\sigma:\{1,\ldots,n\}\to[0,1]^n$
does not depend on the particular choice of $\sigma$ when 
$\sigma\in S(n)$ is such that
$x\cdot\sigma\in\Lambda^n$.  It follows
that $\sigma\in S(n)$
monotonizes $x\in\Delta^n$ iff $\sigma$ makes the following diagram commute:
\beq\label{diagram}
\diagram 
\{1,\ldots,n\}&\rTo^{x}&[0,1]\\
\uTo^{\sigma}&\ruTo(2,2)_{x^\Lambda}&\\ 
\{1,\ldots,n\}&&\\
\enddiagram
\eeq
The inequalities (\ref{eq:monosym}) can now be
restated without explicit reference to $\sigma$.
\begin{proposition}\label{orderwithoutsigma} 
For $x,y\in\Delta^n$, we have $x\sqsubseteq y$ iff 
\ben
\item There exists at least one $\sigma\in S(n)$
such that
$x\cdot\sigma,\,y\cdot\sigma\in\Lambda^n$; 
\item For all $\,i\in\{1,\ldots,n-1\}$ we have $\,x^\Lambda_i\cdot
y^\Lambda_{i+1}\leq x^\Lambda_{i+1}\cdot
y^\Lambda_i\,$.
\een 
\end{proposition} 
\begin{remark}
When $x^\Lambda_{i+1}\not=0\not=y^\Lambda_{i+1}$ the inequalities
express ratios:
\[
{x^\Lambda_i/x^\Lambda_{i+1}}\,\leq\,{y^\Lambda_i/y^\Lambda_{i+1}}\,.
\]
\end{remark} 
Let $x\in\Delta^n$.  Let ${n^x}$ be the cardinality of ${\rm spec}(x)$;
let $x^{\rm
spec}$ be the decreasingly ordered spectrum of $x$. Denote the
multiplicity of value $x_j^{\rm spec}$ in the list $x^\Lambda$ by $n_j^x$,
or, $n_j$ when it is clear from the
context to which state this number applies.  Then, 
set $K_1^{(x)}:=\{1,\ldots,
n_1\}$ and set:
\ben
\item
$\forall j\in\{1,\ldots,n^x\}:\bar{n}_j:=\sum_{i=1}^{i=j}n_i$
\item
$\forall j\in\{2,\ldots,n^x\}:K_j^{(x)}:=\{\bar{n}_{j-1}^{(x)}+1,\ldots,
\bar{n}_j^{(x)}\}$
\een
that is 
$i\in K_j\ \Leftrightarrow\ x^\Lambda_i=x_j^{\rm spec}$. 
The diagram in eq.(\ref{diagram}) then splits up in 
\[
\diagram 
\{1,\ldots,n\}&\rTo^{x}&[0,1]\\
\uTo^{\sigma}&\ruTo(2,2)_{x^{\rm spec}(1)}&\\ 
K_1&&\\
\enddiagram
\quad\quad
\ldots
\quad\quad
\diagram 
\{1,\ldots,n\}&\rTo^{x}&[0,1]\\
\uTo^{\sigma}&\ruTo(2,2)_{x^{\rm spec}(n)}&\\ 
K_{{n^x}}&&\\
\enddiagram
\]
where $x^{\rm spec}(1),\ldots,x^{\rm spec}(n)$ are constant maps.
Requiring commutation then imposes an ordered
partition $(\sigma[K_1],\ldots,\sigma[K_{{n^x}}])$ on $\{1,\ldots,n\}$.  

For $i,j\in\{1,\ldots, n\}$ set $i\sim j$ whenever
$x_i=x_j$. The corresponding equivalence classes then admit a total
ordering $I_1^{(x)}\succ\ldots\succ
I_{{n^x}}^{(x)}$ which is such that $I_k\succ I_l$ whenever for $i\in I_k$
and $j\in I_l$ we have $x_i> x_j$. 
Thus 
\beq\label{eq:degenerationsets}
i\in I_j\ \Leftrightarrow\ x_i=x_j^{\rm spec}\,.
\eeq
The cardinality of $I_j$ is the same as that of $K_j$, namely $n_j$. 
\begin{lemma}\label{Setmonotonization}
For $x\in\Delta^n$ and $\sigma\in S(n)$ we have $x\cdot\sigma\in\Lambda^n$
iff
\[
\forall j\in\{1, \ldots, {n^x}\}:\sigma[K_j]=I_j\,.
\]
\end{lemma}
\noindent{\bf Proof.} 
%%%%%%%%%%%%%%%%
%If $x\cdot\sigma\in\Lambda^n$ the we have 
%$$
%\sigma(i)\in \sigma[K_j]\Leftrightarrow i
%\in K_j\Leftrightarrow x_j^{\rm
%spec}=x_i^\Lambda=(x\cdot\sigma)(i)=x_{\sigma(i)}
%\Leftrightarrow\sigma(i)\in I_j\,.
%$$
%The converse follows by construction.
%%%%%%%%%%%%%%%%
Since by diagram (\ref{diagram}) we have
$x\cdot\sigma\in\Lambda^n\Leftrightarrow \forall
i\in\{1,\ldots,n\}:x_i^\Lambda=(x\cdot\sigma)(i)$ the
equivalence  follows from 
$\sigma(i)\in \sigma[K_j]\Leftrightarrow i\in K_j\Leftrightarrow x_j^{\rm
spec}=x_i^\Lambda$ and 
$\sigma(i)\in I_j\Leftrightarrow x_j^{\rm
spec}=x_{\sigma(i)}=(x\cdot\sigma)(i)$.
\hfill\endproof
\begin{proposition}\label{ListSpecRep}
Each $x\in \Delta^n$ is faithfully represented by the pair 
\ben
\item The ordered partition  ${\cal I}^x:=(I_1,\ldots, I_{{n^x}})$ on
$\,\{1,\ldots,n\}$\,;
\item The $[0,1]$-valued $\,n^x$-element set $\,{\rm spec}(x)$.
\een
Conversely, each such pair defines a state $x\in \Delta^n$ iff
$\,\sum_{j=1}^{j=n^x}n_j\cdot x_j^{\rm spec}=1$.
\end{proposition} 
\noindent{\bf Proof.} 
Direction $\Rightarrow$ of eq.(\ref{eq:degenerationsets}) fixes $x$ given
${\rm spec}(x)$ and
$(I_1,\ldots, I_{{n^x}})$.  The converse follows by construction.
\hfill\endproof\newline

The degeneration of the spectrum of $x\in\Delta^n$ which is now encoded in
the ordered partition ${\cal I}^x$
is of crucial importance w.r.t.~$\sqsubseteq\,$. 
\begin{lemma}[Degeneration] {\rm\cite{CoeMar}}\label{degeneration}
If $x\sqsubseteq y$ in $\Delta^n$, then 
\[
x_i=0\ \Rightarrow \ y_i=0
\quad\quad\&\quad\quad
y_i=y_j>0\ \Rightarrow\ x_i=x_j
\]
\end{lemma}
Thus, degeneration admits a hierarchy in $(\Delta^n,\sqsubseteq)$:

\vspace{1mm}

\begin{center}
\begin{tabular}{|c|} 
\hline
zero-values/degeneration\\
\hline
non-degenerated non-zero values\\ 
\hline
degenerated non-zero values\\ 
\hline
\end{tabular}
\end{center}

\vspace{0mm}\par\noindent
Setting
\[
\hspace{-2mm}\left\{
\begin{array}{lcl} 
n_{0}^{(x)}:={n^x}&\ \ &0\not\in {\rm spec}(x)\\
n_{0}^{(x)}:={n^x}-1\ ,\  \bar{n}_0:=\sum_{i=1}^{i=n_0}n_i \ ,\
I_0:=I_{{n^x}}\
,\ K_0:=K_{{n^x}}&\ \ &0\in {\rm spec}(x)  
\end{array}
\right.
\]
we can express the Degeneration Lemma in terms of ${\cal I}^x$.  
\begin{lemma}[Degeneration${}^{\bf bis}$]\label{degenerationbis} 
If $x\sqsubseteq y$ in $\Delta^n$, then
\[
I_0^x\subseteq I^y_0\quad\quad \& \quad\quad \forall i\in\{1,\ldots, n_0^y\},\exists
j\in\{1,\ldots,
n_0^x\}:I_i^y\subseteq I_j^x\,.
\]
\end{lemma}

\section{Coordinates} 

\begin{definition} 
{\bf(Coordinates)} Let $\,{\rm Coord(\Delta^n)}$ be all $x\in \Delta^n$ with an at most
binary spectrum.  Let the {\it degenerated coordinates\,} $\,{\rm
Ir_\bot(\Delta^n)}$ be the set of all $x\in{\rm
Coord(\Delta^n)}$ with $\,0\in{\rm spec}(x)$.  For $x\in{\rm
Ir_\bot(\Delta^n)}$ let the $x$-{\it axis\,} be 
the set of all $y\in{\rm Coord(\Delta^n)}$ with $I_1^y=I_1^x$ (and thus
also $I_2^y=I_2^x$).
\end{definition}  
As shown in \cite{CoeMar} \S 4.3, ${\rm Ir_\bot(\Delta^n)}$ constitutes a
subposet of $\Delta^n$ which, when
top and bottom are added to it, is
isomorphic to the powerset ${\cal P}(\{1,\ldots,n\})$.  The illustrations
below expose ${\rm
Ir_\bot(\Delta^n)}\cup\{\bot\}$  in the ``triangle''
$\Delta^3$ and the ``tetrahedron''
$\Delta^4$. The figures on the right are their Hasse diagrams.

\medskip\noindent
\begin{minipage}[b]{1.02\linewidth} 
\centering\epsfig{figure=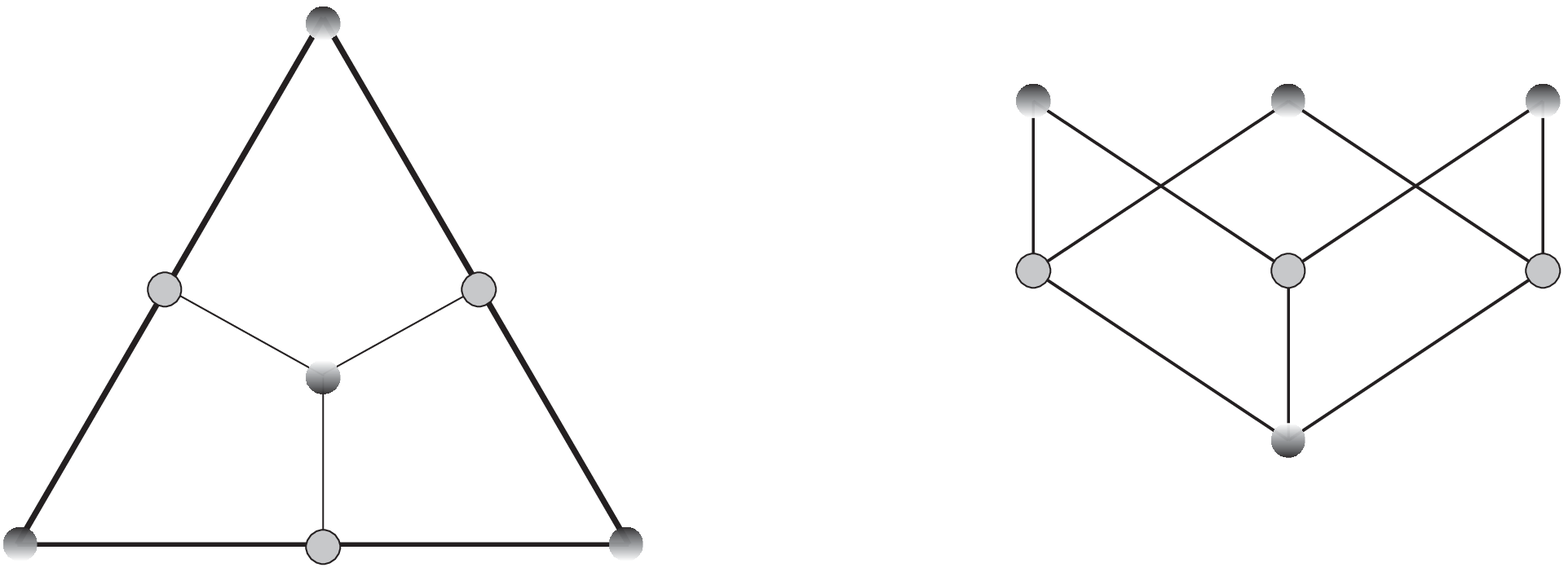,width=285pt}
\end{minipage}\hfill

\medskip\noindent
\begin{minipage}[b]{0.99\linewidth} 
\centering\epsfig{figure=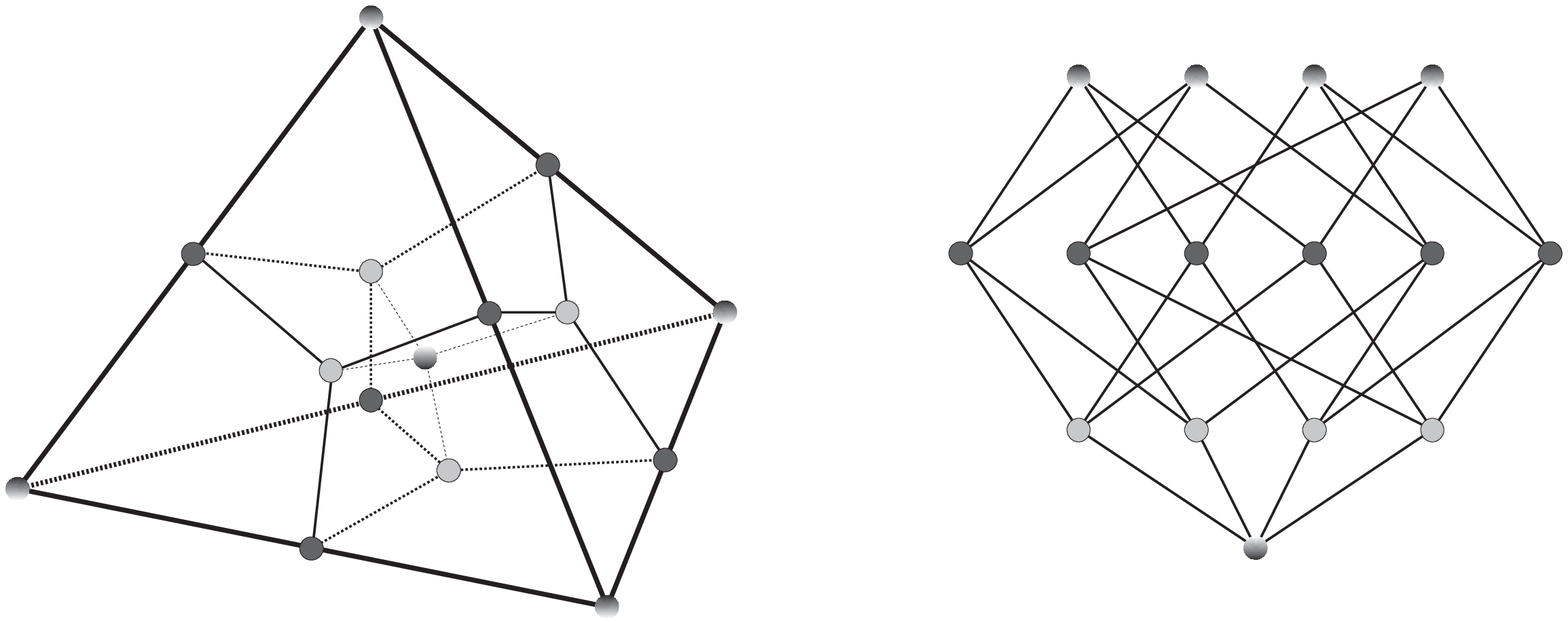,width=295pt}
\end{minipage}\hfill

\smallskip\noindent
The segments represent increase of the order and coincide on the left and
the right, the increase being 
respectively radially and upwardly.  The coordinate axes of 
$\Delta^3$ and 
$\Delta^4$ look as follows. 

\par\vspace{-0.5mm}
\par\noindent
\begin{minipage}[b]{1\linewidth} 
\centering\epsfig{figure=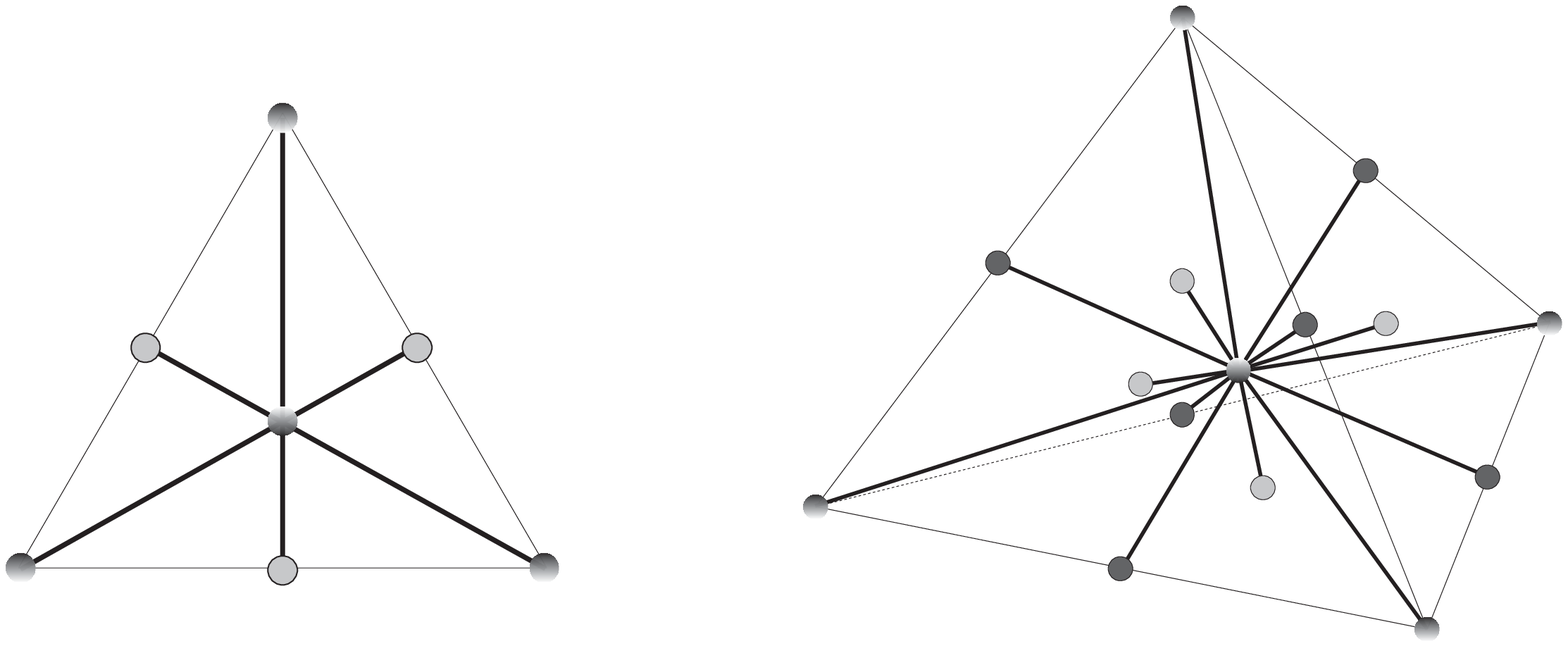,width=280pt}
\end{minipage}\hfill

\begin{proposition}\label{coordareorder}
Coordinates and coodinate axes are order-theoretical:
\bit
\item ${\rm Ir_\bot(\Delta^n)}\cup\{\bot\}$ are the infima of sets in
$\,{\cal
P}({\rm Max}(\Delta^n))\setminus\{\emptyset\}$.  
\item If $x\in{\rm Coord}_{\rm Ir}(\Delta^n):=
{\rm Coord(\Delta^n)}\setminus{\rm Ir_\bot(\Delta^n)}$ then
$\downarrow\!x$ is a chain. 
Conversely, if
$\downarrow\!x$ is a chain then $x\in{\rm Coord(\Delta^n)}$.
\item A coordinate axis is the completion of a maximal 
$\,{\rm Coord}_{\rm Ir}(\Delta^n)$-chain.
\eit 
\par\vspace{-0.5mm}\par
\end{proposition}
\noindent{\bf Proof.} 
Maximal elements and bottom are order-theoretical by definition and so are
all $x\in {\rm Ir}_\bot(\Delta^n)$ 
since by \cite{CoeMar} \S 4.3 we have
$x=\bigwedge\bigl(\uparrow\!x\cap{\rm Max}(\Delta^n)\bigr)$.

For $x\in{\rm Coord(\Delta^n)}\setminus{\rm Ir_\bot(\Delta^n)}$ we have
$x=\bot$ or ${\cal I}^x=(I_1^x,I_2^x)$.
Let $x\not=\bot$.  If $y\sqsubseteq x$ by Lemma
\ref{degenerationbis}  we have $I_1^x\subseteq I^y$ and $I_2^x\subseteq
J^y$ for some $I^y,J^y\in{\cal I}^y$.  Thus, ${\cal I}^y={\cal I}^x$ or $
I_1^y=\{1,\ldots,n\}$.  If $y,z \in\downarrow\!x$ with $y\not=\bot\not=z$
then ${\cal
I}^y={\cal I}^z={\cal I}^x$ and either
$y^+\cdot z^-\leq z^+\cdot y^-$ or $z^+\cdot y^-\leq y^+\cdot z^-$ so $y$
and $z$ compare.
The cases $x=\bot$, $y=\bot$ and $z=\bot$ are trivial so $\downarrow\!x$ 
is a chain.

Let $x\not\in{\rm Coord(\Delta^n)}$.  Then
$\{I_1^x,I_2^x,I_3^x\}\subseteq{\cal
I}^x$. But $y,z\in\Delta^n$ defined by 
\beqa
\begin{array}{ccc}
\ \ {\cal I}^{y}=\left\{I_1^x\ {\bf,}\,\{1,\ldots,n\}\setminus
I_1^x\right\}
&&
\ \ {\cal I}^{z}=\left\{I_1^x\cup I_2^x\ {\bf,}\,\{1,\ldots,n\}\setminus
(I_1^x\cup I_2^x)\right\}\vspace{2mm}
\\
\ \ \ \ y_1^{\rm spec}\cdot x_2^{\rm spec}=x_1^{\rm spec}\cdot y_2^{\rm
spec}
&&
\ \ z_1^{\rm spec}\cdot x_3^{\rm spec}=x_2^{\rm spec}\cdot z_2^{\rm spec}
\end{array}  
\eeqa
(cfr.~Proposition \ref{ListSpecRep}) don't compare although
$y,z\sqsubseteq x$ so $\downarrow\!x$ is not a chain.

From the above we also know that for $x\in{\rm Coord}_{\rm Ir}(\Delta^n)$ and
$y\sqsubseteq x$ we have $y\in{\rm Coord(\Delta^n)}$ and in particular
that $y$ belongs to the same axis as $x$.
Thus for 
$y,z\in x$-axis with $z\not=x$ we have that $y\sqsubseteq w\sqsubseteq z$
forces $w\in x$-axis.  Thus 
$x$-axis$\setminus\{x\}$ is a maximal chain in ${\rm Coord}_{\rm Ir}(\Delta^n)$.
By \cite{CoeMar} Proposition 2.16 we then have 
$x=\bigsqcup (x$-axis$\setminus\{x\})$.
\hfill\endproof\newline
%%%%%%%%%%%%%%%%%%%%%%%%%
%%%%%%%%%%%%%%%%%%%%%%%%%
%%%%%%%%%%%%%%%%%%%%%%%%%
%... {\it coordinate axis}, which are, the chains of the form 
%$$
%\bigl[\bot\,{\bf ,}(1/k,\ldots,1/k,0,\ldots,0)\cdot\sigma\bigr]:=
%\bigl\{(a,\ldots,a,b\ldots,b)\cdot\sigma\in\Delta^n\bigm| a\geq b\bigr\}
%$$
%with $\sigma\in S(n)$ and $(1/k,\ldots,1/k,0,\ldots,0)\in\Delta^n$.
%%%%%%%%%%%%%%%%%%%%%%%%%
%%%%%%%%%%%%%%%%%%%%%%%%%
%%%%%%%%%%%%%%%%%%%%%%%%%

To $x\in\Delta^n\setminus\{\bot\}$ we attribute
${\cal C}^x=\{c(1),\ldots,c({n^x}-1)\}\subset{\rm Coord(\Delta^n)}$ as
{\it its coordinates},
where, using 
Proposition \ref{ListSpecRep}, each $c(j)$ is defined by
\[
{\cal I}^{c(j)}=\left\{\bigcup_{i=1}^{i=j} I_i^x\
{\bf,}\,\bigcup_{i=j+1}^{i={n^x}}I_i^x\right\}
\quad\quad\quad
c(j)_1^{\rm spec}\cdot x_{j+1}^{\rm spec}=x_j^{\rm spec}\cdot c(j)_2^{\rm
spec}\,.
\]
Further we set ${\cal C}^\bot=\emptyset$.
If $0\in {\rm spec}(x)$ we set $c_0:=c(n^x-1)\in{\rm Ir_\bot(\Delta^n)}$.
\begin{theorem}[Decomposition in coordinates]\label{prop:decomposition}
States $x\in\,\Delta^n$ and their coordinates ${\cal C}^x$ are in bijective
order-theoretic correspondence:
\[ 
x=\bigsqcup {\cal C}^x\quad\quad{\rm and}\quad\quad{\cal C}^x={\rm
Max}\bigl({\rm
Coord(\Delta^n)}\cap\!\downarrow\!x\bigr)\setminus\{\bot\}\,.
\]
\end{theorem}
\noindent{\bf Proof.} 
We exclude the trivial case $x=\bot$.
Note that by counting we obtain 
\[
K^{c(j)}_1=\bigcup_{i=1}^{i=j} K_i^x
\quad\
K^{c(j)}_2=\bigcup_{i=j+1}^{i={n^x}}K_i^x
\quad\
K^{c_0}_1=\bigcup_{i=1}^{i={n^x_0}}K_i^x
\quad\
K^{c_0}_2=K_0^x\,.
\]
Let $x\cdot\sigma_x\in\Lambda^n$. By Lemma \ref{Setmonotonization} we have 
$\forall i\in\{1, \ldots, {n^x}\}$ that $\sigma[K_i]=I_i$ and as such we
have 
for all $j\in\{1, \ldots, {n^x_0}-1\}$
\[ 
\sigma_x\left[K^{c(j)}_1\right]=
\sigma_x\left[\bigcup_{i=1}^{i=j} K_i^x\right]=
\bigcup_{i=1}^{i=j} \sigma_x\left[K_i^x\right]=\bigcup_{i=1}^{i=j}
I_i^x=I^{c(j)}_1\,.
\]
Analogously, $\sigma_x[K^{c(j)}_2]=I^{c(j)}_2$,
$\sigma_x\left[K^{c_0}_1\right]=I^{c_0}_1$ and
$\sigma_x\left[K^{c_0}_2\right]=I^{c_0}_2$. 
Thus, again by Lemma
\ref{Setmonotonization}, for all $c(j)\in {\cal C}^x$ we have
$c(j)\cdot\sigma_x\in\Lambda^n$ so $x$ and
$c(j)$ admit joint monotonization. Again, let $j\in\{1,\ldots,n_0^x-1\}$.
We have:
\ben
\item
$c(j)_{\bar{n}_j^x}^\Lambda/c(j)_{\bar{n}^x_j+1}^\Lambda=
x_{\bar{n}_j^x}^\Lambda/x_{\bar{n}_j^x+1}^\Lambda$\,;
\item $c(j)^\Lambda_i/c(j)^\Lambda_{i+1}=1\leq
x^\Lambda_i/x^\Lambda_{i+1}$ for $i\in
\{1,\ldots, \bar{n}^x_0-1\}\setminus \{\bar{n}_j^x\}$\,; 
\item $c(j)_i^\Lambda\cdot x^\Lambda_{i+1}=c(j)_i^\Lambda\cdot 0\leq
c(j)^\Lambda_{i+1}\cdot x^\Lambda_i$ for $i\in\{\bar{n}^x_0,\ldots,n-1\}$.  
\een 
Thus $c(j)\sqsubseteq x$ by Proposition
\ref{orderwithoutsigma}.  Analogously, in the case that $0\in {\rm
spec}(x)$ we have $c_0\sqsubseteq
x$. Thus, $x$ is an upper bound for ${\cal C}^x$.

Let $z\in\Delta^n$ be such that $\forall c\in {\cal C}^x:c\sqsubseteq z$
and $\sigma_x,\sigma_z\in S(n)$ 
such that $x\cdot\sigma_x\in\Lambda^n$ and 
$z\cdot\sigma_z\in\Lambda^n$.  First we construct $\sigma\in S(n)$ that
monotonizes both $x$ and $z$.
Set
$n_z^x:=\sup\bigl(\{0\}\cup\{j\in \{1,\ldots, {n^x}\}\bigm| K_j^x\cap
K_0^z=\emptyset\}\bigr)$.

Assume $n_z^x\not =0$ (if not, skip this paragraph).
We have for $i\in I_1^{c(1)}=I^x_1$ and for $k\in I_2^{c(1)}=\{1,\ldots,
n\}\setminus I^x_1$ that 
$c(1)_i>c(1)_k\not=0$.  Since $c(1)\sqsubseteq z$ we have by Lemma
\ref{degeneration} that 
$\forall i\in I^x_1\,,\,\forall k\in\{1,\ldots, n\}\setminus I^x_1:z_i>
z_k$ 
so $\sigma_z[K^x_1]=I^x_1$.  

\vspace{0.05cm}
{ 
\begin{center}
\begin{picture}(300,40)   
\put(0,0){\line(1,0){300}} 
\put(0,20){\line(1,0){300}} 
\put(0,40){\line(1,0){300}} 
\put(300,0){\line(0,1){40}}  
%%%%%%%%%%%%%%%
\put(0,0){\line(0,1){40}} 
\put(25,7){$I_1^x$} 
\put(60,0){\line(0,1){40}} 

\put(5,27){$I_1^z$} 
\put(18,20){\line(0,1){20}} 

\put(24,27){$\ldots$} 

\put(42,20){\line(0,1){20}} 
\put(46,27){$I_i^z$} 
%%%%%%%%%%%%%%%
\put(69,7){$\ldots$} 
\put(69,27){$\ldots$} 
%%%%%%%%%%%%%%%
\put(90,0){\line(0,1){40}} 
\put(115,7){$I^x\!\!\!\!${\tiny${}_{n^{x}_z}$}} 
\put(150,0){\line(0,1){40}}  

\put(95,27){$I_j^z$}  
\put(108,20){\line(0,1){20}} 

\put(114,27){$\ldots$} 

\put(132,20){\line(0,1){20}} 
\put(136,27){$I^z\!\!\!${\tiny${}_{n^{z}_x}$}} 
%%%%%%%%%%%%%%%
\put(150,0){\line(0,1){40}} 
\put(183,7){$I^x\!\!\!\!${\tiny${}_{n^{x}_z\!+\!1}$}} 
\put(230,0){\line(0,1){20}} 

\put(151,27){$I^z\!\!\!\!${\tiny${}_{n^{\!z}_{\!x}\!+\!1}$}}   
\put(168,20){\line(0,1){20}} 

\put(174,27){$\ldots$} 

\put(192,20){\line(0,1){20}} 
\put(193,27){$I^z\!\!\!\!${\tiny${}_{n^z_0}$}} 
\put(210,20){\line(0,1){20}}  
%%%%%%%%%%%%%%%
\put(234,7){$\ldots$}  
%%%%%%%%%%%%%%%
\put(250,0){\line(0,1){20}} 
\put(252,07){$I^x\!\!\!\!${\tiny${}_{n^x_0}$}} 
\put(270,0){\line(0,1){20}} 
%%%%%%%%%%%%%%%
\put(250,27){$I_0^z$}  
\put(280,07){$I_0^x$}  
\end{picture} 
\end{center}
\[
\ \ \,\,\uparrow\sigma
\]
\begin{center}
\begin{picture}(300,27)   
\put(0,0){\line(1,0){300}} 
\put(0,20){\line(1,0){300}} 
\put(300,0){\line(0,1){20}}  
\put(0,0){\line(0,1){20}} 
\put(60,0){\line(0,1){20}} 
\put(18,0){\line(0,1){20}} 
\put(24,07){$\ldots$} 
\put(42,0){\line(0,1){20}} 
\put(69,7){$\ldots$} 
\put(69,7){$\ldots$} 
\put(90,0){\line(0,1){20}} 
\put(150,0){\line(0,1){20}}  
\put(108,00){\line(0,1){20}} 
\put(114,7){$\ldots$} 
\put(132,00){\line(0,1){20}} 
\put(150,0){\line(0,1){20}} 
\put(230,0){\line(0,1){20}} 
\put(168,0){\line(0,1){20}} 
\put(174,7){$\ldots$} 
\put(192,0){\line(0,1){20}} 
\put(210,0){\line(0,1){20}}  
\put(234,7){$\ldots$}  
\put(250,0){\line(0,1){20}} 
\put(270,0){\line(0,1){20}} 
\put(-2,-11){$0$}  
\put(298,-10){$n$}   
\put(2,7){$K_1^z$} 
\put(278,07){$K_0^x$}    
\end{picture} 
\end{center}
}
\vspace{0.3cm}   

\noindent
By induction on $j\in\{1,\ldots,n_z^x\}$,
since $c(j)\sqsubseteq z$ we have
\[
\forall l\in \bigcup_{l=1}^{l=j-1}I^x_l\,,\,\forall i\in
I^x_j\,,\,\forall k\in\{1,\ldots,
n\}\setminus \bigcup_{l=1}^{l=j}I^x_l:z_l>z_i> z_k
\]
 so $\sigma_z[K^x_j]=I^x_j$.  
Let $n_x^z$ be such that 
$\bigcup_{j=1}^{j=n_x^z}K_j^z=\bigcup_{j=1}^{j=n_z^x}K_j^x$.  Setting 
\[
\forall j\in\{1, \ldots, n_x^z\}:\sigma[K_j^z]:=\sigma_z[K_j^z]=I_j^z
\]
we also obtain
$\forall j\in\{1, \ldots, n_z^x\}:\sigma[K_j^x]=I_j^x$.

Next we set 
\bit
\item $\forall
j\in\{n^z_x+1,\ldots,n^z_0\}:\sigma[K_j^z]:=\sigma_z[K_j^z]=I_j^z$
\item $\sigma[K_{n_z^x+1}^x\cap K_0^z]:=
\sigma_x[K_{n_z^x+1}^x]\cap\sigma_z[K_0^z]=I_{n_z^x+1}^x\cap I_0^z$
\item $\forall K_j^x\subseteq K^z_0:\sigma[K_j^x]:=\sigma_x[K_j^x]=I_j^x$
\eit
Since $c(n^x_z+1)\sqsubseteq z$ we obtain along the same lines as above
that 
$\sigma[K_{n_z^x+1}^x]=I_{n_z^x+1}^x$
and
$\sigma[K_0^z]=I_0^z$.
Conclusively, $\sigma$ monotonizes both $x$ and $z$.
We now verify the inequalities of Proposition \ref{orderwithoutsigma} in
order to prove that $x\sqsubseteq z$.  
\ben
\item
$x_{\bar{n}_j^x}^\Lambda/x_{\bar{n}_j^x+1}^\Lambda=
c(j)_{\bar{n}_j^x}^\Lambda/c(j)_{\bar{n}_j^x+1}^\Lambda\leq
z_{\bar{n}_j^x}^\Lambda/z_{\bar{n}_j^x+1}^\Lambda$ for $j\in
\{1,\ldots, n^x_0\!-1\}$\,;
\item $x^\Lambda_i/x^\Lambda_{i+1}=1\leq z^\Lambda_i/z^\Lambda_{i+1}$ for
$i\in
\{1,\ldots, \bar{n}^z_0\!-1\}\setminus \{\bar{n}_j^x|\,j\in
\{1,\ldots, n^x_0\!-1\}\}$; 
\item $x_i^\Lambda\cdot z^\Lambda_{i+1}=x_i^\Lambda\cdot 0\leq
x^\Lambda_{i+1}\cdot z^\Lambda_i$ for $i\in\{\bar{n}^z_0,\ldots,n-1\}$.    
\een 
Conversely, ${\cal C}^x={\rm Max}\bigl({\rm
Coord(\Delta^n)}\cap\downarrow\!x\bigr)\setminus\{\bot\}$ follows by Lemma
\ref{degenerationbis} and the
fact that 
$c(j)_1^{\rm spec}\cdot x_{j+1}^{\rm spec}=x_j^{\rm spec}\cdot c(j)_2^{\rm
spec}$ maximizes those coordinates
below $x$ that are on the same axis.  
\hfill\endproof\newline   

One easily verifies that this decomposition is irreducible,
that is, ${\cal C}^x$ is the infimum for inclusion of all finite ${\cal
C}\subseteq{\rm Coord}(\Delta^n)$ with $x=\bigsqcup{\cal C}$.
We proceed by characterizing the sets that
arise as ${\cal C}^x$ for some $x$.  
It will follow that each ${\cal C}^x$ implicitly  
is an ordered list, the order being induced by the order on the
irreducibles that label the axes to which each $c(j)\in{\cal C}^x$ belongs.
\begin{proposition}\label{CharCoorSets}
$\{c(1),\ldots,c(m)\}$ are the coordinates of some $x\in\Delta^n$ iff
\ben
\item $m\leq n-1$
\item $x^1\sqsupset\ldots\sqsupset x^{m}$ where $\forall
j\in\{1,\ldots,m\}:c(j)\in x^j$-axis$\setminus\{\bot\}$
\item $c(j)=x^j\Rightarrow j=m$
\een
\par\vspace{-0.5mm}\par
\end{proposition}
\noindent{\bf Proof.}  
For each $c(j)$ we obtain $x^j$ such that $c(j)\in x^j$-axis by setting
${\cal I}^{x^j}={\cal I}^{c(j)}$ and
$0\in{\rm spec}(x^j)$.  (2.)~is then easily verified. (1.)~and (3.)~are
obvious.  Conversely, defining
${\cal I}^x$ by intersecting the sets ${\cal I}^{c(j)}$ for all
$j\in\{1,\ldots,m\}$ and imposing
$c(j)_1^{\rm spec}\cdot x_{j+1}^{\rm spec}=x_j^{\rm spec}\cdot c(j)_2^{\rm
spec}$ 
we construct $x\in\Delta^n$ which satisfies ${\cal
C}^x=\{c(1),\ldots,c(m)\}$. 
\hfill\endproof\newline   

\section{Isomorphisms}

\begin{theorem}[Isomorphisms]\label{thm:orderiso}
Order-isomorphisms of $\,(\Delta^n,\sqsubseteq)$ are in bijective correspondence with
pairs consisting of 
\bit
\item $\sigma\in S(n)${\rm ,}~~{\rm ($\sim$\,labeling the elements in
${\rm Max}(\Delta^n)$)}
\item $2^n-2$ order-isomorphisms of $[0,1]$.~~{\rm($\sim$\,gauging
each coordinate axis)} 
\eit
\par\vspace{-2.5mm}\par 
\end{theorem} 
\noindent{\bf Proof.} 
Let $h:\Delta^n\to\Delta^n$ be an order-isomorphism.  We have
$h(\bot)=\bot$. Since $h[{\rm Max}(\Delta^n)]={\rm Max}(\Delta^n)$ 
this induces a permutation $\sigma\in
S(n)$ via $\sigma(e_i)=h(e_i)$. This
permutation $\sigma$ extends  to one on all $x\in{\rm Ir_\bot(\Delta^n)}$ since
they are of the form
$\bigwedge\bigl(\uparrow\!x\cap\max(\Delta^n)\bigr)$ which on its turn
extends by Proposition \ref{coordareorder}
to all coordinate axis (as a whole).  For each coordinate axis 
set 
\[
f_x:x{\rm\mbox{-}axis}\to x{\rm\mbox{-}axis}::y\mapsto
h(y\cdot\sigma^{-1})
\]
Since $h$ is an order-isomorphism, so is $f_x$.  The action on
each $x\in\Delta^n$ is then implied by $x=\bigsqcup {\cal C}^x$.
Conversely, let 
$\{f_x:x{\rm\mbox{-}axis}\to x{\rm\mbox{-}axis}\}$ be
the $2^n-2$ order-isomorphisms of $[0,1]$ and let $\sigma\in S(n)$. Define an order
isomorphism
\[
h:\Delta^n\to\Delta^n::y\mapsto\bigsqcup\{f_{x}(c(j))\cdot\sigma\mid
c(j)\in{\cal C}^y, c(j)\in
x{\rm\mbox{-}axis}\}\,.
\]
Existence of the suprema follows from Proposition \ref{CharCoorSets},
bijectivity from Theorem
\ref{prop:decomposition} and monotonicity from ${\cal C}^x={\rm
Max}\bigl({\rm
Coord(\Delta^n)}\cap\!\downarrow\!x\bigr)\setminus\{\bot\}$.  Indeed, when
$x\sqsubseteq y$ then 
this forces each $c(j)\in{\cal C}^x$ to have an upper bound in ${\cal
C}^y$ since then 
$\downarrow\!x\subseteq\downarrow\!y$. Applying this argument to $h^{-1}$
yields strictness.
\hfill\endproof\newline
\begin{corollary}\label{cl:iso} 
The identity is the only order-isomorphism of $\,(\Delta^n,\sqsubseteq)$ which preserves
both
{\rm Max}$(\Delta^n)$ and Shannon entropy (or any other map that is
strictly increasing on coordinate
axis).
\end{corollary}
\noindent{\bf Proof.} 
By Theorem \ref{thm:orderiso} it suffices to verify that Shannon entropy
is strictly increasing on each
coordinate axis. Then its preservation forces all maps 
${\{f_x:x{\rm\mbox{-}axis}\to x{\rm\mbox{-}axis}\mid x\in{\rm
Ir}_\bot(\Delta^n)\}}$ to be identities. 
\hfill\endproof\newline

By definition of $D^n$ there exists an order-isomorphism
$h:D^n\to(\Delta^n,\sqsubseteq)$.  
A map $\mu:D^n\to[0,1]$ is induced by commutation of  
\[
\diagram 
{[0,1]}&\rTo^{id}&{[0,1]}\\
\uTo^{\mu}&&\uTo_{\mu}\\ 
D^n&\rTo^{h}&(\Delta^n,\sqsubseteq)\\
\enddiagram
\]
Corollary \ref{cl:iso} implies
that if $\mu:D^n\to[0,1]$ is fixed, no other 
order-isomorphism $h':D^n\to(\Delta^n,\sqsubseteq)$ satisfying
$\forall i:h(e_i)=h'(e_i)$ makes the diagram commute.
Thus, the pair $(D^n, \mu:D^n\to[0,1])$ defines a unique
gauge $h:D^n\to \Delta^n$ which assigns to each 
$x\in D^n$ a unique list of numbers $h(x)\in\Delta^n$.

\section{Probability from logic}\label{sec:const}  

We will reconstruct $D^n$  from $A:={\rm P}(\{1,\ldots,n\})$ in
order-theoretic manner.

\medskip\noindent
{\bf Formal procedure}.  Let $A$ be a bounded poset. Let $\Gamma$ be a
bounded chain.\footnote{The construction and Proposition
\ref{pr:ConstPoset} still hold for $\Gamma$
any bounded poset.}
\ben
\item Denote by $A_{0,1}^*$ the poset obtained by removing the top and
bottom
from $A$ and by reversing the order.
\item Let ${\rm MChain}(A_{0,1}^*)$ be all maximal chains
$\vec{a}=\{a_1\sqsupset\ldots\sqsupset
a_{n-1}\}$ in $A_{0,1}^*$.
In benefit of lucidity we assume that all these chains
have length $n-1$.\footnote{The construction and Proposition
\ref{pr:ConstPoset} still
hold without this assumption.}
\item Denote by ${\rm Cl}_\top(\Gamma^{n-1})$ the set of all
$\Gamma$-valued tuples
$\vec{\gamma}=(\gamma_1,\ldots,\gamma_{n-1})$ subjected to the
closure\footnote{${\rm Cl}_\top$ indeed acts as a closure operator on the
pointwisely ordered complete lattice
$\Gamma^{n-1}$, and thus, ${\rm
Cl}_\top(\Gamma^{n-1})$ is
itself a complete lattice. For all $n\geq 2$ monotone states constitute
complete lattices since $(\Lambda^n,\sqsubseteq)\cong{\rm
Cl}_\top([0,1]^{n-1})$. 
Moreover, $(\Delta^n,\sqsubseteq)$ admits arbitrary non-empty
infima and any subset of $\Delta^n$ with an upper bound has a
supremum  w.r.t.~$\sqsubseteq\,$~\cite{Lattices}.}
\[
\forall i<j\in\{1,\ldots, n-1\}:\gamma_i=\top\ \Rightarrow\
\gamma_j=\top\,.
\]
\item Set
$[A^*_{0,1},\Gamma]:=\{\vec{a}\cdot\vec{\gamma}\mid\vec{a}\in{\rm
MChain}(A_{0,1}^*)\,,\vec{\gamma}\in{\rm Cl}_\top(\Gamma^{n-1})\}$\,.
\item Introduce the pointwisely induced relation
\[
\vec{a}\cdot\vec{\gamma}\sqsubseteq\vec{b}\cdot\vec{\varphi}
\ \,\Longleftrightarrow\ \,\vec{a}=\vec{b}\ {\rm and}\ \forall
i\in\{1,\ldots,
n-1\}:\gamma_i\sqsubseteq\varphi_i\,.
\]
\item Define the indices:
\[
I(\vec{\gamma}):=\{i\in\{1,\ldots,n-1\}\mid\gamma_i\not\in\{\bot,\top\}\}\,;
\]
\[
\iota(\vec{\gamma}):=\inf\{i\in\{1,\ldots,n-1\}\mid\gamma_i=\top\}\,.
\]
Let $\overline{[A^*_{0,1},\Gamma]}$ be the set of equivalence classes in
$[A^*_{0,1},\Gamma]$ obtained for
\[
\vec{a}\cdot\vec{\gamma}=\vec{b}\cdot\vec{\varphi}
\ \,\Longleftrightarrow\ \, \vec{\gamma}=\vec{\varphi}\ {\rm and}\
(i\in I(\vec{\gamma})\cup\{\iota(\vec{\gamma})\}\Rightarrow a_i=b_i)\,.
\]
\item
Finally, $\overline{[A^*_{0,1},\Gamma]}$ inherits the relation
$\,\sqsubseteq\,$ on $[A^*_{0,1},\Gamma]$,
explicitly,
\[
\vec{a}\cdot\vec{\gamma}\sqsubseteq\vec{a}\cdot\vec{\varphi}\ \,
\Longrightarrow\
\,[\vec{a}\cdot\vec{\gamma}]\sqsubseteq[\vec{a}\cdot\vec{\varphi}]\,.
\]
\een
\begin{proposition}\label{pr:ConstPoset}
$\Bigl(\overline{[A^*_{0,1},\Gamma]}\,{\bf ,}\sqsubseteq\Bigr)$ is a poset
with a bottom.
\end{proposition}
\noindent{\bf Proof.}
We have to prove anti-symmetry and transitivity of $\,\sqsubseteq\,$ on
$\overline{[A^*_{0,1},\Gamma]}$.

{\it Anti-symmetry}. Let
$\vec{a}\cdot\vec{\gamma}\sqsubseteq\vec{a}\cdot\vec{\varphi}$ and
$\vec{b}\cdot\vec{\gamma}\sqsupseteq\vec{b}\cdot\vec{\varphi}$ with
$[\vec{a}\cdot\vec{\gamma}]=[\vec{b}\cdot\vec{\gamma}]$ and
$[\vec{a}\cdot\vec{\varphi}]=[\vec{b}\cdot\vec{\varphi}]$.  We must then
for all $i\in\{1,\dots, n-1\}$
both have $\gamma_i\sqsubseteq\varphi_i$ and
$\varphi_i\sqsubseteq\gamma_i$ from which
$\vec{a}\cdot\vec{\gamma}=\vec{a}\cdot\vec{\varphi}$ and thus
$[\vec{a}\cdot\vec{\gamma}]=[\vec{a}\cdot\vec{\varphi}]$ follows.

{\it Transitivity}.  Let
$\vec{a}\cdot\vec{\gamma}^-\sqsubseteq\vec{a}\cdot\vec{\gamma}$ and
$\vec{b}\cdot\vec{\gamma}\sqsubseteq\vec{b}\cdot\vec{\gamma}^+$ with
$[\vec{a}\cdot\vec{\gamma}]=[\vec{b}\cdot\vec{\gamma}]$. We have to prove
that
$[\vec{a}\cdot\vec{\gamma}^-]\sqsubseteq[\vec{b}\cdot\vec{\gamma}^+]$.  We
define $\vec{c}\in{\rm
MChain}(A_{0,1}^*)$ as follows.
For $i\in I(\vec{\gamma}):c_i:=a_i=b_i$, for
$i\in\{\iota(\vec{\gamma}),\ldots,n-1\}:c_i:=a_i$ and in all other cases,
that is $\gamma_i=\bot$, we set
$c_i:=b_i$. Since $\gamma_i^-\sqsubseteq\gamma_i$ implies
$\gamma_i=\bot\Rightarrow\gamma_i^-=\bot$ and
$\gamma_i\sqsubseteq\gamma_i^+$ implies
$\gamma_i=\top\Rightarrow\gamma_i^+=\top$
it respectively follows that
$[\vec{c}\cdot\vec{\gamma}^-]=[\vec{a}\cdot\vec{\gamma}^-]$ and
$[\vec{c}\cdot\vec{\gamma}^+]=[\vec{a}\cdot\vec{\gamma}^+]$.
Thus,  since
$\vec{c}\cdot\vec{\gamma}^-\sqsubseteq\vec{c}\cdot\vec{\gamma}^+$
due to $\gamma_i^-\sqsubseteq\gamma_i\sqsubseteq\gamma_i^+$ for all
$i\in\{1,\dots, n-1\}$ we obtain
$[\vec{a}\cdot\vec{\gamma}^-]\sqsubseteq[\vec{b}\cdot\vec{\gamma}^+]$.

Finally, choosing $\vec{a}$ arbitrary in ${\rm MChain}(A_{0,1}^*)$ and
setting
$\vec{\gamma}=(\bot,\ldots,\bot)$, we obtain $[\vec{a}\cdot\vec{\gamma}]$
as the bottom of
$\overline{[A^*_{0,1},\Gamma]}$.
\hfill\endproof
\begin{problem}
A categorical variant of this construction would be desirable.
\end{problem}
\begin{lemma}\label{S(n)cong}
${\rm MChain}({\rm P}(\{1,\ldots,n\})_{0,1}^*)\cong S(n)$ as sets.
\end{lemma}
\noindent{\bf Proof.}
The sets ${\rm MChain}({\rm P}(\{1,\ldots,n\})_{0,1}^*)$ and $S(n)$ are in
bijective correspondence via
$\forall i\in\{1,\ldots,n-1\}:a_i=\bigvee\bigl\{e_j\bigm|
j\in\sigma\bigl[\{1,\ldots, i\}\bigr]\bigr\}$\,.
\hfill\endproof\newline
\begin{theorem}[Construction of classical states] \label{th:Conctclass}
Let $n\geq 2$.
\[
\biggl(\overline{\left[{\rm P}(\{1,\ldots,n\})^*_{0,1}\,{\bf
,}\,[0,1]\right]}\,{\bf
,}\sqsubseteq\biggr)\cong\bigl(\Delta^n,\sqsubseteq\bigr)
\]
\end{theorem}
\noindent{\bf Proof.}
Assume $\xi:[0,1]\to[1,\infty]$ to be
a fixed order isomorphism. Let $\vec{a}\cdot\vec{\gamma}\in[{\rm
P}(\{1,\ldots,n\})^*_{0,1}\,{\bf
,}\,[0,1]]$. We can define a set ${\cal C}^{\vec{a}\cdot\vec{\gamma}}$ of
coordinates as follows.
For each $a_i\in\vec{a}$ such that
$i\in I(\vec{\gamma})\cup\{\iota(\vec{\gamma})\}$ define $c(i)\in{\rm
Coord}(\Delta^n)$ such that
${\cal I}^{c(i)}=(I^i,\{1,\ldots, n\}\setminus I^i)$ where $I^i$ is
implicitly defined by
$a_i=\bigvee\{e_j\mid j\in I^i\}$, and by setting
$c_1^i/c_2^i=\xi(\gamma_i)$ whenever $\gamma_i\not=1$ and
$c_2^i=0$ otherwise. The set ${\cal
C}^{\vec{a}\cdot\vec{\gamma}}=\{c^i\mid i\in
I(\vec{\gamma})\cup\{\iota(\vec{\gamma})\}\}$ satisfies the conditions in
Proposition \ref{CharCoorSets} and as
such ${\cal C}^{\vec{a}\cdot\vec{\gamma}}={\cal C}^x$ for
$x=\bigsqcup{\cal
C}^{\vec{a}\cdot\vec{\gamma}}$.  For
$\vec{a}\cdot\vec{\gamma},\vec{b}\cdot\vec{\varphi}\in[{\rm
P}(\{1,\ldots,n\})^*_{0,1}\,{\bf ,}[0,1]]$ we have ${\cal
C}^{\vec{a}\cdot\vec{\gamma}}={\cal
C}^{\vec{b}\cdot\vec{\varphi}}$ iff
$\vec{a}\cdot\vec{\gamma}\sim\vec{b}\cdot\vec{\varphi}$ in the above
defined equivalence relation on
$[{\rm P}(\{1,\ldots,n\})^*_{0,1}\,{\bf
,}[0,1]]$.
Due to uniqueness of the decomposition in coordinates (Theorem
\ref{prop:decomposition}) we obtain an injective
correspondence  between
$\overline{[{\rm P}(\{1,\ldots,n\})^*_{0,1}\,{\bf ,}\,[0,1]]}$ and
$\Delta^n$ and by Proposition \ref{CharCoorSets} it follows that it is
also surjective.

We now show that this correspondence also preserves the order. It follows
from the definition of
$\,\sqsubseteq\,$ that for
$[\vec{a}\cdot\vec{\gamma}], [\vec{b}\cdot\vec{\varphi}]\in\overline{[{\rm
P}(\{1,\ldots,n\})^*_{0,1}\,{\bf
,}\,[0,1]]}$ we have
$[\vec{a}\cdot\vec{\gamma}]\sqsubseteq[\vec{b}\cdot\vec{\varphi}]$ iff
there exists $\vec{c}\in{\rm
MChain}({\rm P}(\{1,\ldots,n\})_{0,1}^*)$ such that
$\vec{c}\cdot\vec{\gamma}\in[\vec{a}\cdot\vec{\gamma}]$ and
$\vec{c}\cdot\vec{\varphi}\in[\vec{b}\cdot\vec{\varphi}]$ and such that
$\vec{c}\cdot\vec{\varphi}\sqsubseteq\vec{b}\cdot\vec{\varphi}$. 
Moreover,
\ben
\item 
Existence of $\vec{c}\in{\rm
MChain}({\rm P}(\{1,\ldots,n\})_{0,1}^*)$ with
$\vec{c}\cdot\vec{\gamma}\in[\vec{a}\cdot\vec{\gamma}]$ and
$\vec{c}\cdot\vec{\varphi}\in[\vec{b}\cdot\vec{\varphi}]$ coincides with
existence of $\sigma\in S(n)$ which
monotonizes both
$x=\bigsqcup{\cal C}^{\vec{a}\cdot\vec{\gamma}}$ and $y=\bigsqcup{\cal
C}^{\vec{b}\cdot\vec{\varphi}}$, extending the isomorphism in Lemma \ref{S(n)cong}.
\item Due to $c_1^i/c_2^i=\xi(\gamma_i)$
for $\gamma_i\not=1$ and
$c_2^i=0$ for $\gamma_i=1$, the pointwisely defined order for
$\vec{\gamma}$ and $\vec{\varphi}$ induces
eq.(\ref{eq:monosym}) for $x=\bigsqcup{\cal C}^{\vec{a}\cdot\vec{\gamma}}$
and $y=\bigsqcup{\cal
C}^{\vec{b}\cdot\vec{\varphi}}$.
\een
Explicit verification of the above completes the proof.
\hfill\endproof\newline
\newpage
\begin{remark}
It should be clear to the reader that the metric on $[0,1]$ doesn't
play any role, i.e.,~$[0,1]$ should be read as an
order-theoretic abstraction.
\end{remark}
\begin{remark}\label{AltConst}
The alternative representation of classical states in Proposition
\ref{ListSpecRep} incarnates as an instance of an alternative
formulation of this 
construction. It simplifies the definition of the set
$\overline{[A^*_{0,1},\Gamma]}$ but one looses lucidity
w.r.t.~the pointwise nature of the induced order.  Explicitly,
let ${\rm Chain}(A_{0,1}^*)$ be all chains in $A_{0,1}^*$, let 
$\Gamma_{\bot,\top}:=\Gamma\setminus\{\bot,\top\}$, let 
$\Gamma_\bot:=\Gamma\setminus\{\bot\}$, let
\[
{\rm Cl}_\top(\Gamma^{n-1}_\bot):=
   \left\{(\gamma_1,\ldots,\gamma_{k})\mid
        k\leq n-1; 
        \gamma_1,\ldots,\gamma_{k-1}\in\Gamma_{\bot,\top};
        \gamma_k\in\Gamma_\bot 
   \right\}\,,
\]
and denoting by $|-|$ the length of a list we obtain
\[
\overline{[A^*_{0,1},\Gamma]}\cong
   \left\{\vec{a}\cdot\vec{\gamma}\mid
        \vec{a}\in{\rm Chain}(A_{0,1}^*)\,;
        \vec{\gamma}\in{\rm Cl}_\top(\Gamma^{n-1}_\bot)\,;
        |\vec{a}|=|\vec{\gamma}|
   \right\}\,.
\]
\end{remark}
\begin{theorem}[Construction of quantum states] \label{th:Conctquant} Let
$n\geq 2$.
\[
\biggl(\overline{\left[(\mathbb{L}^n)^*_{0,1}\,{\bf
,}\,[0,1]\right]}\,{\bf
,}\sqsubseteq\biggr)\cong
\bigl(\Omega^n,\sqsubseteq\bigr)\,.
\]
\end{theorem}
We omit the proof here. We do want to expose a remarkable fact. Contrary
to a Boolean
algebra where orthogonality is captured by the order via
\[
a\perp b\Leftrightarrow a\wedge b=0\,,
\]
the lattice $\mathbb{L}^n$ admits many different
orthocomplementations.\footnote{An
orthocomplementation on a lattice $L$ is an antitone involution $(-)':L\to
L$ which satisfies $a\wedge
a'=0$ and $a\vee a'=1$. It provides an orthogonality relation via
$a\perp b\Leftrightarrow a\leq b'$.} Mixed quantum states, due to the
particular status measurements have in quantum theory, are measures
$\omega:\mathbb{L}^n\to[0,1]$ which satisfy
\beq\label{Orthadd}
a\perp b\ \Rightarrow\ \omega(a\vee b)=\omega(a)+\omega(b)\,.
\eeq
By Gleason's theorem \cite{Gleason} these are in bijective correspondence
with the density matrices
(the set which we denoted in \cite{CoeMar} by $\Omega^n$).  We can
envision a constructor
$\underline{\rm Val}[-]$\,, acting on all posets $D$ that go equipped with
an orthogonality relation
$\perp$\,, which assigns to each $(D,\perp)$ the (monotone) measures
$\omega:D\to[0,1]$ that satisfy
(\ref{Orthadd}), ordered along the lines of
\cite{CoeMar}.\footnote{Besides domain-theoretic differences,
a sharp distinction between $(\Delta^n,\sqsubseteq)$ and the
Jones--Plotkin probabilistic powerdomain
\cite{Plotkin} is the fact that the Bayesian order is a relation on
probability measures {\it contra\,}
the Jones--Plotkin construction which builds a probabilistic universe on
top of a pre-existing
order-theoretic structure; we claim that the epistemic nature of
probability has a primal
mathematical structure on its own which is order-theoretic.}
We have
\[
\underline{\rm Val}\left[\bigl({\rm
P}(\{1,\ldots,n\}),(-)^c\bigr)\right]\cong(\Delta^n,\sqsubseteq)
\quad\&\quad
\underline{\rm
Val}\left[\bigl(\mathbb{L}^n,(-)'\bigr)\right]\cong(\Omega^n,\sqsubseteq)\,,
\]
with $(-)^c$ the Boolean complement and $(-)'$ any orthocomplementation on
$\mathbb{L}^n$.
The above {\it entropic geometry construction\,} however enables to
produce an isomorphic copy
of $(\Omega^n,\sqsubseteq)$ without the requirement of specification of an
orthocomplementation on
$\mathbb{L}^n$.  Indeed, we obtain the constructor
${\underline{\rm EntGeom}[-]}$ which acts on any poset and satisfies
\[
\underline{\rm EntGeom}\left[{\rm
P}(\{1,\ldots,n\})\right]\cong(\Delta^n,\sqsubseteq)
\quad\&\quad
\underline{\rm
EntGeom}\left[\mathbb{L}^n\right]\cong(\Omega^n,\sqsubseteq)\,.
\]
A detailed exposition and elaboration on this matter is in preparation \cite{CoeMar2}.

As a third example let $D$ be a $(n+1)$-element chain with
$n\geq 2$. Then
\[
\biggl(\overline{\left[D^*_{0,1}\,{\bf,}\,[0,1]\right]}\,{\bf
,}\sqsubseteq\biggr)\cong
\bigl(\Lambda^n,\sqsubseteq\bigr)\,.
\]
This {\bf construction of monotone states} constitutes a 
fragment of both the classical and the
quantum states construction; it constitutes the {\it atom\,} of the
entropic geometry construction.
 
\medskip\noindent
{\bf Interpretation}. 
The Boolean logic $A\cong{\rm P}(\{1,\ldots,n\})$ can be generated by introducing
disjunction on its atomic properties $\{e_1,\ldots,e_n\}$. These atomic
properties provide {\it total specification\,} of the system.
A disjunction $e_i\vee\ldots\vee e_j$ only provides {\it partial
specification\,} of the system. It however still provides {\it total knowledge\,}
on truth of the property $e_i\vee\ldots\vee e_j$.
We could emphasize this by writing $(e_i\vee\ldots\vee e_j,\top)$ 
standing for ``total knowledge on truth of $e_i\vee\ldots\vee e_j$''.    

Rather than only providing total knowledge on properties, 
we can increase expressiveness
by making {\it partiality of
knowledge\,} explicit: We will write ${(e_i\vee\ldots\vee e_j,\gamma)}$ with
$\gamma\in\Gamma_{\bot,\top}$ the degree of partiality of our
knowledge. This for example allows to refine $(e_i\vee e_j,\top)$ to 
$((e_i,\gamma),(e_i\vee e_j,\top))$ standing for ``most likely the
state of the system is $e_i$, with certainty it is either $e_i$ or
$e_j$, and the degree to which it is rather in $e_i$ than in $e_j$ is
$\gamma$''.  The list 
\[
((a_1:=e_i,\gamma_1),\ldots,
(a_{k-1},\gamma_{k-1}),
(a_k:=a_{k-1}\vee e_j,\top))
\]
with $\gamma_1,\ldots,\gamma_{k-1}\in\Gamma_{\bot,\top}$
then expresses that {\it most likely\,} the system is in pure state $e_i$,
with certainty it is either in one of the states that span $a_k$, and the 
degree to which $a_i$ is more likely than $a_{i+1}$ is encoded as
$\gamma_i$; any occurence of $(a_j,\bot)$ should be conceived as a {\it
void\,} statement --- their explicit ommitance exactly provides the alternative
construction of Remark \ref{AltConst}; we can extend the
list with a
superfluous tail, or, if it has lenght $n$, delete $(1,\top)$ from it, in order 
to obtain a maximal chain $\vec{a}=(a_1, \ldots,
a_{n-1})$. Such a list provides full
specification of our knowledge about the system.
This explains why we can reproduce all classical states by means of
this construction.

An order relation arises naturally.  We compare
$\vec{a}\cdot\vec{\gamma}$ and $\vec{a}\cdot\vec{\varphi}$ 
by pointwisely comparing $\vec{\gamma}$ and $\vec{\varphi}$;
we have $\vec{a}\cdot\vec{\gamma}\sqsubseteq\vec{a}\cdot\vec{\varphi}$
iff each property in $\vec{a}$ is less likely to be
true for $\vec{a}\cdot\vec{\gamma}$ than it is 
for $\vec{a}\cdot\vec{\varphi}$.  The void statements then cause an
equivalence relation on the set of all possible specifications of this
kind.

Note that we do {\it not\,}
have to require $i\leq j\Rightarrow \gamma_i\sqsubseteq\gamma_j$ since
$\gamma_i,\gamma_j\in\Gamma_{\bot,\top}$
encode ratios of decrease of likelyness of
the newly added atomic property in the next list element as compared to the
remaining head of the list; on the other hand whenever $i\leq j$ then
${\gamma_i=\top}\Rightarrow \gamma_j=\top$ 
has to be fulfilled since in that case we have $a_i\Rightarrow
a_j$. The bounds $\bot$ and $\top$ indeed play a distinct role in the
construction, one is void and the other captures truth. 

This reasoning also extends to chains in arbitrary posets when
envisioned as algebras of properties of a system:
Whenever we have $(a_i,\gamma_i)$ with $\gamma_i\not=\top$, we add a
weaker property $a_{i+1}\in A$ which is such that $a_i\Rightarrow a_{i+1}$, untill
we obtain $a_k$ such that $(a_k,\top)$ --- this $a_k$ can of course be
$1$. The construction of quantum states illustrates this claim.

\medskip
\noindent
{\bf The geometric picture}.  We illustrate the above for the case of
$n=3$. 
  
\bigskip\noindent
\begin{minipage}[b]{1\linewidth} 
\hspace{1.1cm}\begin{picture}(70,80)
\put(0,48){\LARGE $A\, =\ $}
\end{picture}
\begin{picture}(120,113)   
\put(52.5,112){\scriptsize $1$}
\put(-25,75){\scriptsize $e_1\!\vee\!e_2$}
\put(29,75){\scriptsize $e_1\!\vee\!e_3$}
\put(112,75){\scriptsize $e_2\!\vee\!e_3$}
\put(-9,32){\scriptsize $e_1$}
\put(45,32){\scriptsize $e_2$}
\put(112,32){\scriptsize $e_3$}
\put(52.5,-8){\scriptsize $0$}
\centering\epsfig{figure=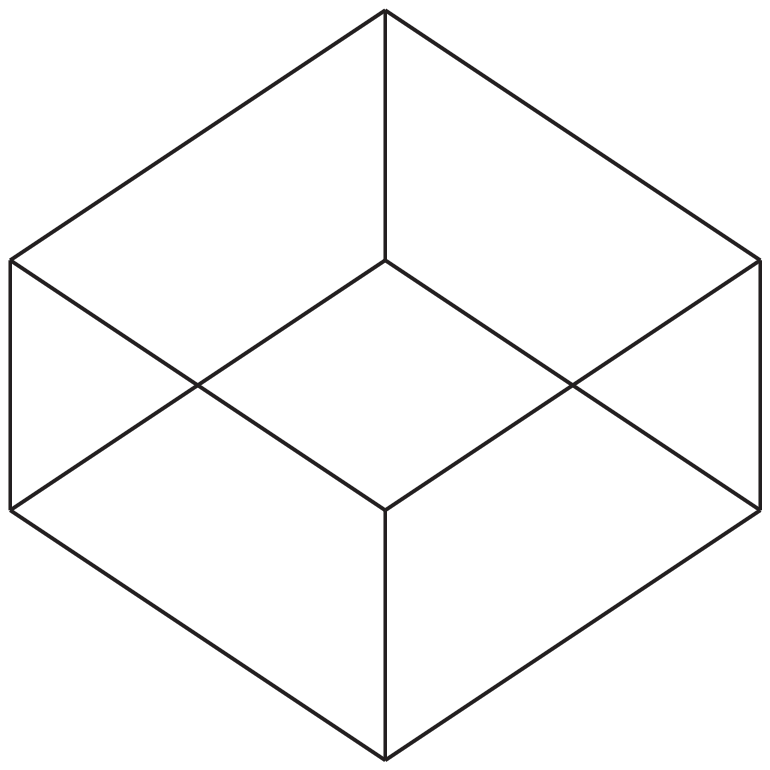,width=127pt}
\end{picture}
\end{minipage}\hfill

\par\vspace{-0.0cm}\par
 
\medskip\noindent
\begin{minipage}[b]{1\linewidth} 
\hspace{5.2cm}
\begin{picture}(70,80)
\put(0,35){\LARGE $\!\!\!\!\!A^*_{0,1} =\ \ \, $}
\end{picture}
\begin{picture}(120,80)   
\put(-25,34){\scriptsize $e_1\!\vee\!e_2$}
\put(29,34){\scriptsize $e_1\!\vee\!e_3$}
\put(112,34){\scriptsize $e_2\!\vee\!e_3$}
\put(-10,80){\scriptsize $e_1$}
\put(44,80){\scriptsize $e_2$}
\put(112,80){\scriptsize $e_3$}
\hspace{-2.4mm}
\centering{\hspace{0.2mm}\epsfig{figure=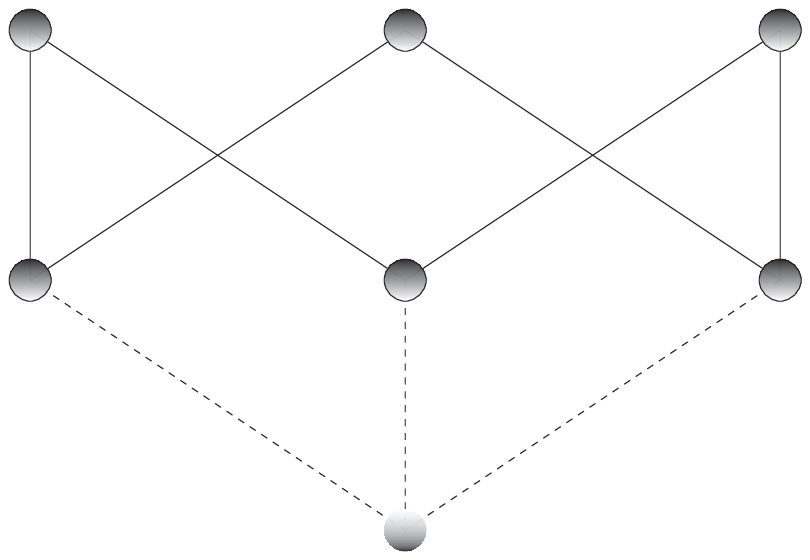,width=130pt}}
\end{picture}
\end{minipage}\hfill

\noindent
Pairing elements of $A^*_{0,1}$ with those of $\Gamma$ creates increasing
``lines'' which all rise
from a common source, namely the ``void'' statement (denoted as $\bot$).

\bigskip\noindent
\begin{minipage}[b]{1\linewidth} 
\hspace{4.1cm}
\begin{picture}(120,80)   
\put(-25,34){\scriptsize $e_1\!\vee\!e_2$}
\put(29,34){\scriptsize $e_1\!\vee\!e_3$}
\put(112,34){\scriptsize $e_2\!\vee\!e_3$}
\put(-6,80){\scriptsize $e_1$}
\put(48,80){\scriptsize $e_2$}
\put(116,80){\scriptsize $e_3$}
\put(76,-8){\scriptsize $\bot$}
\hspace{-2.5mm}
\centering{\hspace{0.2mm}\epsfig{figure=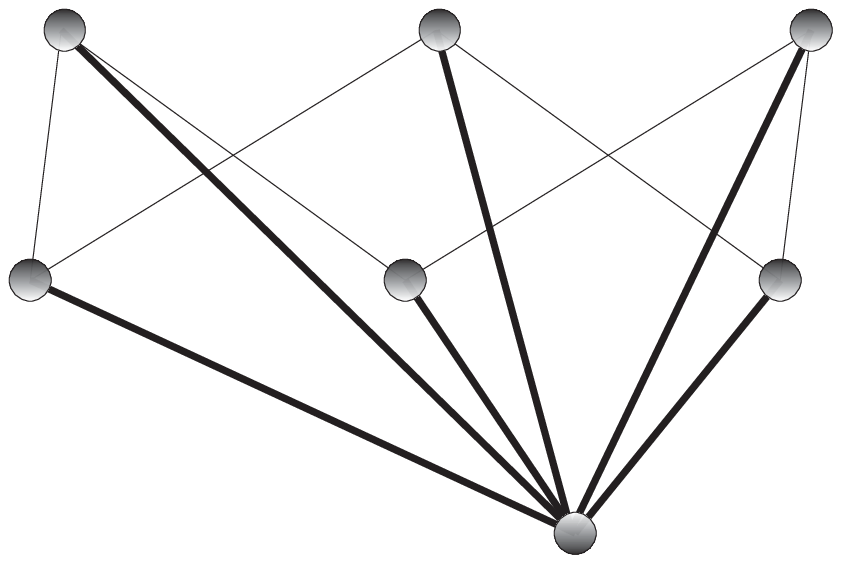,width=129.3pt}}
\end{picture}
\end{minipage}\hfill

\bigskip\noindent
Finally, the formation of lists for all chains in ${\rm
MChain}(A_{0,1}^*)$ fills the regions enclosed by the
corresponding lines resulting in a triangle.

\bigskip\noindent
\begin{minipage}[b]{1\linewidth} 
\hspace{1.2cm}\begin{picture}(120,82)    
\put(-6,80){\scriptsize $e_1$}
\put(48,80){\scriptsize $e_2$}
\put(116,80){\scriptsize $e_3$}
\hspace{-2.4mm}
\centering\epsfig{figure=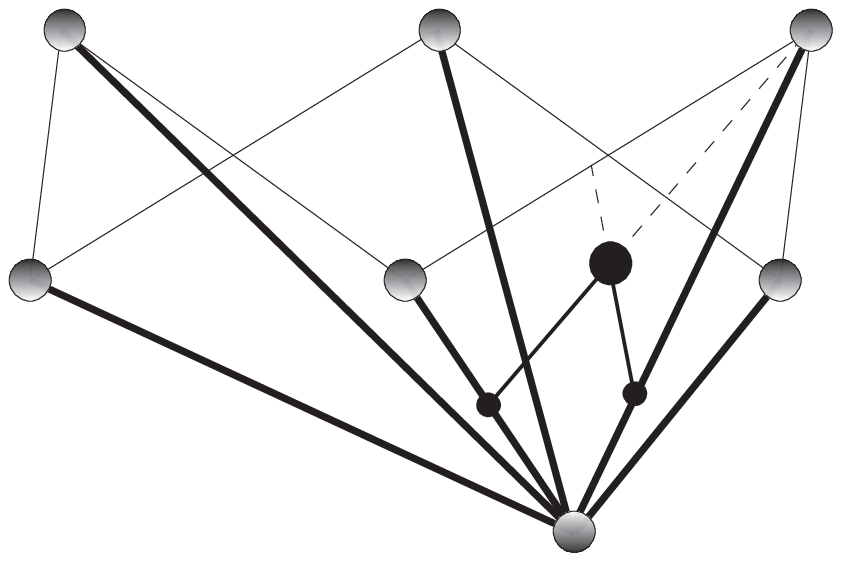,width=129.3pt}     
\end{picture}
\quad\quad
\begin{picture}(10,80)
\put(0,40){\huge $\cong$}
\end{picture}
\quad\quad
\begin{picture}(120,100)   
\put(-6,-2){\scriptsize $e_1$}
\put(54,103){\scriptsize $e_2$}
\put(115,-2){\scriptsize $e_3$}
\centering\epsfig{figure=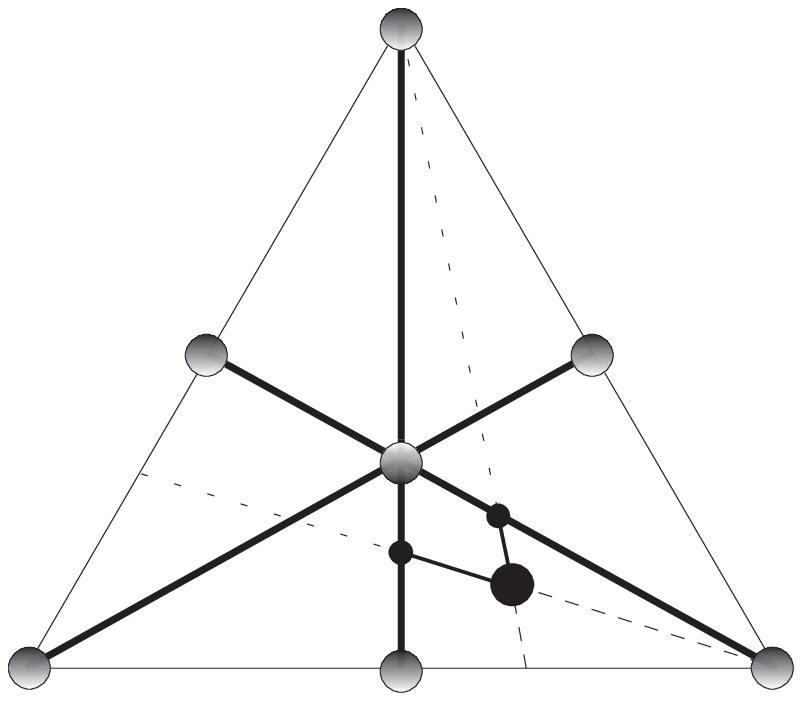,width=114.3pt} 
\end{picture}
\end{minipage}\hfill
  
\medskip\noindent
Note how the formation of lists of pairs (=\,conjunctive) corresponds with the
generation of points as
joins of coordinates ($\sim$\,reversed order).   

Entropic geometry is not merely a geometry of lines but one of
directed lines. 
The triangle or the tetrahedron are not merely convex geometric objects.  For
example, the center of the triangle is a special point from which directed
lines emerge, which stand for the decrease of entropy.  
In a dynamic perspective 
where the lines $\Gamma$ obtain the connotation of {\it flow}, the bounds
$\bot$ and $\top$ obtain the connotation of {\it initiation\,} and {\it
termination}.
The fact that the 4-tuple $(A,\Gamma, \bot,\top)$ generates an entropic
geometry by the above presented
systematic formal procedure can then be interpreted as
\begin{center}
Entropic Geometry = Logic + Flow + Initiation + Termination\,.
\end{center}

\section{Acknowledgements} 

The phrase ``Entropic Geometry'' arose in exchanges with
Keye Martin. I thank 
him for discussing the content and presentation of this paper.  
I thank Samson A.~and Prakash P.~for
logistic, Dusko P.~for recreational and Rhada J.~for
gastronomic support, and for their constructive feedback on \cite{CoeMar}. 
All three referees provided constructive comments.

\end{document}